\documentclass[a4paper]{aa}
\def \th {\thinspace}

\def \degmark{^\circ}

\def\approxgt{\mathrel{\hbox{\rlap{\lower.55ex \hbox {$\sim$}}
  \kern-.3em \raise.4ex \hbox{$>$}}}}
\def\approxlt{\mathrel{\hbox{\rlap{\lower.55ex \hbox {$\sim$}}
  \kern-.3em \raise.4ex \hbox{$<$}}}}
\def \ref {\reference{}}

\def \sun {\hbox {$\odot$}}
\def \degmark{^\circ}

\usepackage{graphicx}
\begin{document}


\title{A comparison of neutron star blackbody luminosities in LMXB
with the theory of accretion flow spreading on the stellar surface}

\author{M. J. Church\inst{1,2}
\and N. A. Inogamov\inst{3}
\and M. Ba\l uci\'nska-Church\inst{1,2}}

\offprints{M. J. Church}

\institute{School of Physics and Astronomy, University of Birmingham,
           Birmingham B15 2TT\\
           email: mjc@star.sr.bham.ac.uk, mbc@star.sr.bham.ac.uk
\and 
           Astronomical Observatory, Jagiellonian University, ul. Orla 171, 30-244 Cracow, Poland
\and
           Landau Institute for Theoretical Physics, Russian Academy of Science,\\
           Kosygin Street 2, V-334, 117940 Moscow, Russian Federation\\
           email: nail@landau.ac.ru}
                    

\date{Received 26 June 2001; Accepted 11 April 2002}
\authorrunning{Church et al.}
\titlerunning{Blackbody emission in LMXB}

\abstract{
We present a comparison of the results of the {\it ASCA} survey of LMXB
with the Inogamov \& Sunyaev theory of accretion flow spreading on the
surface of neutron stars.
The {\it ASCA} survey of LMXB of Church \& Ba\l uci\'nska-Church (2001)
revealed a systematic variation of the luminosity of blackbody emission
from the neutron star spanning 3 decades in total X-ray luminosity
suggesting that the level of blackbody emission is controlled by the physics 
of the inner disk/stellar interface, which we can hope to understand.
Two types of explanation exist: firstly that there is radial flow
between the inner disk and star at all vertical positions above the
orbital plane so that the height of the disk directly
determines the area of star emitting. Secondly, the height of the
emitting region on the star is not directly related to the disk 
properties but depends on the mass accretion rate as suggested by
Inogamov \& Sunyaev (1999) in their theory of accretion flow spreading
on the stellar surface. We find that the survey results for the
emitting area agree with this theory at the lowest luminosities. However,
for higher luminosities, the blackbody emission is 
stronger than predicted by spreading theory suggesting that the
emitting area is controlled by radial flow between disk and star.
}
\maketitle

\section{Introduction}

The nature of the X-ray emission in Low Mass X-ray Binaries (LMXB)
has for many years been controversial. It is clear that
the accretion flow deposits its energy partly in the accretion disk
resulting in heating and X-ray emission, and partly in a boundary
layer at the surface of the star or on the stellar surface itself,
but the nature of the emission has not been agreed. Firstly,
there has been the question of how many continuum components there are.
The spectra of the less bright LMXB are clearly dominated by
Comptonization which led to the use of the Generalized Thermal model
$E^{-\Gamma}\cdot {\rm e}^{-E/E_{\rm CO}}$ in the fitting of {\it Exosat} 
spectra (White et al. 1988). Only in the case of brighter
sources did spectral fitting demand that a blackbody component was
added to this. Mitsuda et al. (1989) proposed a model consisting of
multi-colour disk (MCD) blackbody emission from the inner accretion disk  
plus blackbody emission from the neutron star, Comptonized in a 
region local to the star, and this two-component model is able to fit
many sources well. Secondly, there is the identification of the
blackbody component in LMXB spectra. Opinion has been divided between
location on the surface or boundary layer at the star, or in the inner
accretion disk, and this controversy continues to recent times.
There are also physical differences between models involving 
X-ray emission in the boundary layer and those involving the 
stellar surface (see below). In the former case, it is assumed that 
the angular momentum in the inner disk adjusts to that of the star 
in a boundary layer in the disk which
becomes the source of X-rays. Thirdly, the location, geometry and size
of the Comptonizing region have been controversial. Many theoreticians 
have favoured Comptonization taking place in an inner, hot region
close to the neutron star.

However, the dipping class of LMXB provides answers to these problems
although this has not been widely appreciated. This group of $\sim $10
sources exhibit decreases in X-ray intensity at the orbital period
due to absorption in the bulge in the outer disk where the accretion
flow from the companion star impacts. Emission models are more strongly 
constrained in these sources, as models have to fit not only the
spectrum of the quiescent source as in other types of LMXB, but also several 
levels of dipping. Spectral evolution is not simple, and cannot be
fitted by absorbed one-component models. It has been shown by
analysis of data from many observations with {\it Exosat}, {\it ASCA},
{\it BeppoSAX} and {\it Rossi-XTE} that
these sources are well-described by a model consisting of
point-like blackbody emission plus Comptonized emission from an
extended accretion disk corona (ADC) (e.g. Church et al. 1997, 1998; 
Ba\l uci\'nska-Church et al. 1999, 2000, 2001; Smale et al. 2001). 
The blackbody is  absorbed
rapidly in dipping showing that it is point-like, and has temperatures
between 0.9 and $\sim $2 keV. The Comptonized emission is shown by
spectral analysis to be removed gradually, and the complex spectral
evolution is well-described by progressive covering of this component
in dipping by the absorber, the covering fraction rising smoothly in
many cases from zero to unity. This, of course, requires the
Comptonizing region to be extended allowing the size to be estimated.
Moreover, measurement of dip ingress times directly provides the size
of the region, and values of the radius of this region for several
sources typically of 50,000 km have been determined (Church 2001).
It is clear that a very extended Comptonizing region with high electron
temperature can only be identified with an ADC. The dipping sources
also indicate that this must be a {\it thin corona}, since if
an ADC of this large radius was spherical, it is unlikely that the
absorber on the outer disk would be able to overlap the source region
completely to produce the 100\% deep dipping observed in many sources
(Smale et al. 2001). It is likely that the ADC in all dipping LMXB
is very extended, which can probably be related to the strong effects
of the neutron star forming the ADC by evaporation of the accretion disk.
Moreover, it is not expected that the dipping sources are atypical, 
and so an extended ADC is expected in all LMXB. The large measured size
of the ADC has several significant consequences as discussed below.

Firstly, these measurements of the ADC radius
$r_{\rm {ADC}}$ allow us to rule out Comptonization models that
involve a localized region in the neighbourhood of the neutron star.
Secondly, the ADC size measurements have an important
consequence for blackbody emission observed from LMXB. 
It is expected that a substantial fraction of the total available energy
will be dissipated initially in the accretion disk, implying a large
disk blackbody component. However, because 
the ADC is very extended, all of the hot inner regions of the
disk will be covered by ADC, and consequently it is expected,
given the high optical depth of the ADC (Church 2001),
that all disk blackbody emission will become Comptonized in the ADC
so that no blackbody emission from the disk itself should be visible.
It is also clear on the basis of the thin, flat disk and thin, flat ADC
geometry, that thermal emission from the disk must be the major source
of seed photons for Comptonization in the ADC.
Thus, blackbody emission clearly identified in the spectra of many
LMXB (e.g. Church \& Ba\l ucinska-Church 2001, see below) must be
identified with the neutron star/boundary layer without any
contribution from the disk being expected. This contrasts with
a recent tendency to fit a disk blackbody plus {\it comptt} Comptonization model
routinely to {\it BeppoSAX} data (e.g. Guainazzi et al. 1998; Sidoli 
et al. 2001; Oosterbroek et al. 2001). This is thus inconsistent 
with the implications of the large measured sizes of Accretion Disk Coronae.

The large ADC size is also relevant to the {\it comptt} Comptonization model used 
in the above fitting.
It has been claimed that this is preferable at low energies ($<$ 1 keV) 
to a cut-off power law representation of Comptonization as
the latter implies that the spectral flux density continues to increase
with decreasing photon energy, even for very low energies below 0.1 keV.
Clearly, at very low energies the flux must decrease because of the lack of seed photons.
The {\it comptt} model due to Titarchuk (1994) 
calculates a Comptonized spectrum assuming a sea of soft thermal photons and so
avoids an increase of flux at the lowest energies. It
uses the Wien approximation to the seed photon spectrum,
valid for $h\nu $ $>>$ $kT$. However, usage of this model in which values of $kT$ from spectral fitting
as high as 1 keV and 2 keV are derived (e.g. Guainazzi et al. 1998) will
contravene the assumptions of the model, and will overestimate 
the seed photon spectrum, and thus the Comptonized spectrum by as much 
as 100\% at 1 keV. Thus a model with $kT$ = 2 keV will only be valid above 10
keV. Regarding the availibility of seed photons, we can calculate whether there is any shortage
of seed photons below 1 keV as follows. 
The large size of the ADC referred to above 
ensures that seed photons produced in the accretion disk
out to a radius of $\sim $50,000 km will be Comptonized as the disk
is covered by ADC out to this radius. 
We have calculated the multi-colour disk blackbody spectrum of 
this part of the disk  
assuming a thin disk temperature profile $T(r)$ by integration 
in radius (for appropriate values of mass accretion rate).
It is clearly dominated by the soft photons from outer radii. For
a luminosity of $\rm {10^{36}}$ erg s$^{-1}$,
the photon spectral flux density forms a broad peak between
0.001--0.01 keV, and for a luminosity of $\rm {10^{38}}$ erg s$^{-1}$,
the flux peaks between 0.01--0.1 keV, showing 
that there is no deficiency of low energy photons.
Thus, the cut-off power law is a perfectly good description
of the Comptonized spectrum to energies at least as low as 0.1 keV, whereas application
of the {\it comptt} model without restricting $kT$ for the seed photons to
suitably low values will be invalid.

Finally, the extended size of the ADC has implications for reflection 
in LMXB. Because all of the inner disk is covered by ADC, illumination
of the disk by the neutron star will not be possible, and ``hot''
reflection of this source from the ADC will take place. This will
not however, have the absorption features of a reflection component
from the accretion disk. Moreover, although the ADC itself can
illuminate the disk and lead to reflection, the high values of optical
depth obtained for the ADC (Church 2001) suggest that this component
will be reprocessed in the ADC and so not observed. This may explain the
lack of detections of reflected components in LMXB. In a black hole
binary such as Cyg\th X-1, the ADC radius is much smaller (Church 2001),
and so a reflection component can be expected, as observed.

\subsection{The {\it ASCA} survey}
The above two-component model used to explain dipping LMXB spectra 
has been applied recently to a survey of all classes of LMXB, 
including the Atoll and Z-track sources (Church \& Ba\l ucinska-Church 2001), 
and found to fit all sources well. As part of this survey, it was shown that 
the blackbody emission cannot originate in the accretion disk, since fitting a two-component
model consisting of multi-colour disk blackbody plus a Comptonization term gave in many 
cases a value for the inner disk radius $r_{\rm i}$ more than
10 times smaller than the radius of the neutron star assumed to be $\sim$ 10 km.
Thus the blackbody
emission has to be from the neutron star and these results support
the idea that disk blackbody emission is fully Comptonized in the ADC.
The main result of the survey was the systematic behaviour that it
revealed in the neutron star blackbody emission as reproduced in Fig. 1.
In brighter sources, the 1--30 keV blackbody luminosity $L_{\rm {BB}}$ 
approached 50\% of the total luminosity (the dotted line) as expected in the Newtonian
approximation. In fainter sources, $L_{\rm {BB}}$ fell to much smaller 
fractions of the total. It was demonstrated that, as the total luminosity 
increased, the blackbody emitting area increased by a factor of 5400 across the
sample, whereas the blackbody temperature $kT_{\rm {BB}}$ changed (in fact, 
decreased) by 40\% (i.e. a factor of 0.60), corresponding to a change by a 
factor of 8 in $T^{\rm 4}$. Thus, the area is the most important factor in
determining the level of blackbody emission.

\begin{figure}[!h] 
\begin{center}                                            
\includegraphics[width=66mm,angle=270]{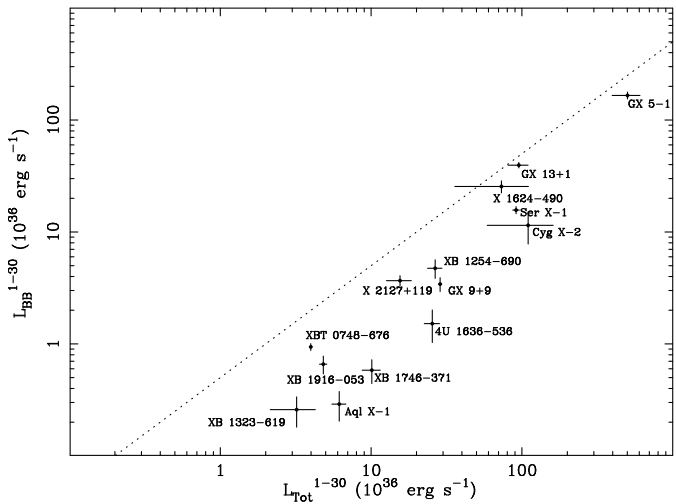}                     
\caption{Variation of blackbody luminosity in the band 1--30 keV with
total luminosity in the same band from the {\it ASCA} survey of LMXB
of Church \& Ba\l uci\'nska-Church (2001)}
\end{center}
\end{figure}

\subsection{Possible modification of the blackbody}

The survey assumed that there was simple blackbody emission from the
surface of the neutron star, and we next discuss this assumption.
Firstly, modification of the blackbody was proposed by Rutledge et al. (1999) for
transiently-accreting LMXB in quiescence, having $kT$ typically 0.2 keV
and luminosity $L$ $\sim$ 10$^{33}$ erg s$^{-1}$. The modification arises
from higher energy photons escaping from greater depths in the atmosphere.
This is clearly inapplicable to LMXB not in quiescence emitting at luminosities of 
10$^{36}$--10$^{38}$ erg s$^{-1}$ (Rutledge priv comm).

Secondly, although in the study of X-ray bursts there is observational
evidence that there is {\it sometimes} modification of the blackbody by electron 
scattering in the neutron star atmosphere during bursts, there is no such
evidence in the case of non-burst emission. Fitting a simple blackbody
in a case where modification did take place would lead to 
errors in $kT$ and the emitting area.
There are two pieces of evidence for modification in bursts 
(see the detailed discussion in Ba\l uci\'nska-Church et al. 2001).
Firstly, values of $kT$ of $\sim$3 keV have been obtained for a fraction of 
bursts from fitting a simple blackbody to the peak, which implies super-Eddington 
emission. Secondly, an increase in blackbody radius $R_{\rm BB}$
while the blackbody temperature 
decreased during burst decay was found in the source XB\th 1636-536 
(Inoue et al. 1984; Sztajno et al. 1985). The increase of $R_{\rm BB}$ was 
thought to be unreal and a consequence of fitting a simple blackbody when 
the emission was modified blackbody. However, whether this effect is observed
depends on whether spectral fitting allows for blackbody emission by the 
non-bursting source (Sztajno et al. 1986), and so is not strong evidence for 
modification. More recently, Kuulkers et al. (2002) have presented evidence
that during the cooling phase of burst decay in GX\th 17+2, the
blackbody is {\it not} modified.

In fact, whether modification takes place depends
critically on the electron density $n_{\rm e}$ which determines whether 
absorption processes or electron scattering dominate the opacity. Various 
theoretical descriptions of the neutron star atmosphere in bursting have been 
given, by Fujimoto et al. (1981), Paczy\'nski (1983), London et al. 
(1984, 1986), Ebisuzaki et al. (1984), Ebisuzaki (1987) and Madej (1991).
However, there are very large disparities between the electron density 
values in these works, $n_{\rm e}$ varying between $10^{\rm 24}$-- $10^{32}$
cm$^{-3}$. While an atmosphere with $n_{\rm e}$ = $10^{\rm 24}$ cm$^{-3}$ is 
clearly dominated by electron scattering, a higher density atmosphere is not.
Thus the theoretical picture is unclear and we have to rely on 
observational evidence that some bursts only have a modified 
spectrum. For these, the effect of modification on measured values
of $kT$ depends on the ratio of colour temperature to effective temperature 
$T_{\rm col}$/T$_{\rm eff}$ for which recent theoretical values are low,
i.e. 1.1 in the work of Madej \& R\'o\.za\'nska (2000) (Czerny, priv comm), 
compared with earlier values which were higher. Values can be derived using
the London et al. model of $\sim $1.6 for $kT_{\rm eff}$ = 2 keV appropriate 
to a burst peak, $\sim $1.3 for 1.5 keV, and $\sim $1.1 for 1 keV (assuming
hydrogen composition). Thus even these factors imply relatively small corrections
for non-burst emission. The effect on emitting area
depends on this ratio squared, so in terms of recent theory can reach only 20\%.
In non-burst emission, there is actually no evidence for modification
at all and so it would be inappropriate to apply a correction factor.
On the contrary, the {\it ASCA} survey results in which the blackbody
radius approaches 10 km in bright sources, i.e. consistent with the neutron 
star radius provide evidence that there is no large modification. 
Thus it is unlikely that the results of the present paper would be affected.

\begin{figure}[!h]                                             
\begin{center}
\includegraphics[width=66mm,angle=270]{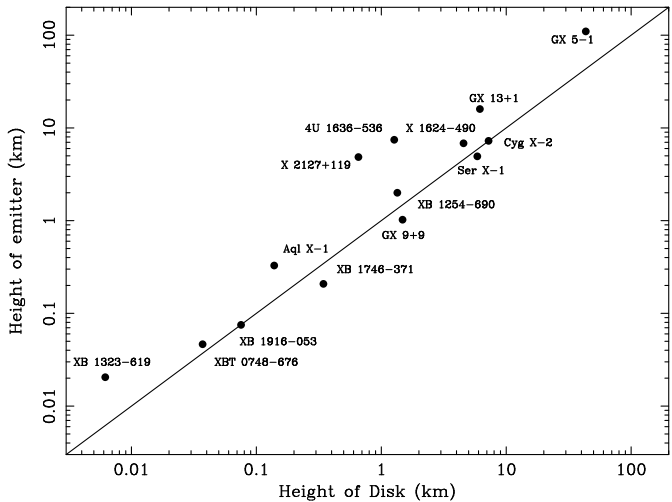}                     
\caption{Variation of the height $h$ of the blackbody emission region
on the surface of the neutron star with the height of the inner,
radiatively-supported accretion disk calculated at the radial position
where $p_{\rm {r}}$ = 10 $p_{\rm {g}}$}
\end{center}
\end{figure}

In Fig. 2, the {\it ASCA} survey results are replotted
in the form of half-height $h$ of the emitting region on the neutron star
assumed to be an equatorial strip, more specifically, a sphere
intersected by two parallel planes, having
emitting area 4$\,\pi\,
R\, h$ where $R$ is the radius of the neutron star. $h$ varied
between 20 m (at the lowest luminosities) and the full radius 
of the star $\sim$ 10 km. These heights are (for almost all of the data)
substantially larger than the half-height of the inner accretion disk
calculated using thin disk theory (Shakura \& Sunyaev 1976)
at some position close to the surface of the neutron star (e.g. 1.001$R$).
This thin disk half-height is $\sim $20 m for a luminosity of 
$\rm {10^{36}}$ erg s$^{\rm -1}$
and $\sim $40 m for $\rm {10^{38}}$ erg s$^{\rm -1}$.
However, the inner disk is expected to be radiatively supported over 
a wide range of luminosities $>$ $\rm {3\times 10^{36}}$ erg s$^{\rm -1}$
and consequently has much larger vertical extent. 
The radiatively-supported disk merges at
some radius with the thin disk and the half-height $H(r_{\rm {10}})$ was 
calculated at the radius at which the radiation pressure was 10 times the gas 
pressure (Czerny \& Elvis 1987). Remarkably, there was good agreement
between $h$ and $H(r_{\rm {10}})$ for the majority of the survey sources,
the agreement spanning 3 decades in each quantity as shown in Fig. 2
(Fig. 3, Church \& Ba\l uci\'nska-Church 2001). 
This agreement suggests that the properties of the inner disk
determine in some way the blackbody emitting area. In the case of the
bright source GX\th 5-1, the agreement is poor with $h$ = 110 km and
$H(r_{\rm {10}})$ = 43 km. The blackbody radius $R_{\rm {BB}}$ is 33 km
showing that the data cannot be explained by emission from the neutron
star alone, suggesting that the emitting region has expanded to a
spherical cloud centred on the star. However, there is good agreement
between the height of the cloud at 33 km (= $R_{\rm {BB}}$) plotted
as a triangle in Fig. 2 and $H(r_{\rm {10}})$. A similar point is plotted
for the blackbody radius of the source GX\th 13+1. There is thus some
evidence that, for these points also, $H$ determines $h$.
In the present paper, we compare the survey results with possible
mechanisms leading to the agreement of $h$ = $H$, specifically i) radial flow between the
accretion disk and neutron star, and ii) accretion flow spreading on the surface of the 
neutron star (Inogamov \& Sunyaev 1999).

The agreement between $h$ and $H$ suggests that the properties of the 
inner disk determine the blackbody emitting area in some way. However,
there are at least two possible explanations of the agreement. Firstly, 
there may be radial flow across the gap between the inner disk and 
the star at all vertical positions above the orbital plane so that the
disk height approximately sets the height of the emitting area on the star.
This requires there to be a stable, radiatively-supported inner accretion
disk, although it has been suggested that the radiative disk is 
subject to instabilities. These may not however, disrupt the disk in a
major way as their characteristic timescale is short (see Sect. 4). 
Secondly, there may be spreading of the accretion flow on the neutron star
surface as proposed by Inogamov \& Sunyaev (1999). In this case,
accretion flow between the disk and the star will be essentially in the orbital plane. 
In this mechanism discussed below, the accretion flow
spreads vertically over the neutron star, the height reached depending
on the mass accretion rate and thus on the total luminosity.
The emitting area depends on $L$, and the quantity $H$ is simply 
a measure of $L$; thus the agreement of $h$ with $H$ represents
the variation of emitting area with luminosity (and does not
result from a direct mapping of the disk height onto the star).
We next compare the survey results with this theory.

\section {Accretion flow on the surface of the neutron star}

In the often accepted standard theory of accretion flow interaction with 
neutron stars, it is assumed that the adjustment in angular velocity
between the accreting material with high Keplerian velocity and the
slower rotation of the star takes place in a boundary layer located in
the inner disk (Shakura \& Sunyaev 1973; Lynden-Bell \& Pringle 1974;
Pringle \& Savonije 1979; Papaloizou \& Stanley 1986; Popham et al.
1993; Bisnovatyi-Kogan 1994; Popham \& Narayan 1995). This takes place
{\it via} turbulent friction between differentially rotating layers in
the disk. Thus X-ray emission would take place in this inner disk.

\begin{figure}[!ht]                                             
\begin{center}
\includegraphics[width=66mm,angle=270]{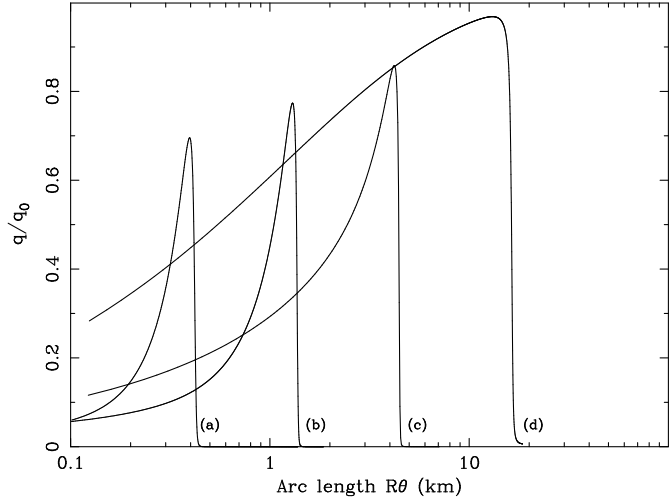}                      
\caption{Variation of X-ray flux $q$ emitted from the neutron star
expressed as a fraction of the Eddington flux $q_{\rm {0}}$
with arc length $R\theta$ for 4 cases in which the source luminosity
is (a) 1\%, (b) 4\%, (c) 20\% and (d) 80\% of the Eddington limit}
\end{center}
\end{figure}

Inogamov \& Sunyaev (1999) recently proposed a radically different
approach in which the adjustment in angular
velocity takes place on the surface of the star, material meeting the
star in the equatorial plane and flowing vertically on the stellar
surface away from the plane and forming a spreading layer.
Viscous interaction with the underlying surface takes place 
while the angular velocity of the flow material differs
from that of the star,
leading to heating and X-ray emission which rises to a
peak at a particular arc distance $R\,\theta$ measured on the surface
of the star from the equator. The emitted flux then falls rapidly with
$R\,\theta$ as energy is radiated.
Thus, for a given luminosity, there is a band on 
the surface of the star where the X-ray emission takes place. 
A system of equations was set up for the spreading layer considered
essentially as a stellar atmosphere, consisting
of hydrostatic equilibrium, radiative equilibrium, an equation of
state containing gas and radiation pressure and the Stefan-Boltzmann
law. The opacity was assumed to be dominated by Thomson scattering.
Transformation and solution of these equations was carried out for a
number of values of accretion rate, i.e. neutron star spreading layer luminosity.
In particular, results were presented for the variation of X-ray flux
$q$ defined as the energy emitted per unit area per second,
with arc distance. Fig. 3 reproduces the results of Inogamov \& Sunyaev for
the distribution of X-ray flux with $R\,\theta$ for 4 values of the
neutron star luminosity equal to 1\%, 4\%, 20\% and 80\% of the Eddington limit
$L_{\rm Edd}$. The flux is given 
relative to the Eddington flux $q_{\rm {0}}$, where

\[L_{\rm {Edd}} \,=\,4\, \pi\, R^{\rm {2}}\,q_{\rm {0}} \,=\, {\rm
1.26\times 10^{38}} \,{M\over M_{\sun}}{\rm erg\; s^{-1}},\] and

\[q_{\rm {0}}\,=\,{G\,M\, m_{\rm {p}}\,c\over R^{\rm {2}}\,\sigma_{\rm {T}}}\]

\noindent
where $M$ is the mass of the neutron star, $m_{\rm {p}}$ is the proton 
mass and $\sigma _{\rm {T}}$ is the Thomson cross section. $L_{\rm {Edd}}$ 
is, of course, independent of the radius assumed for the neutron star. 
Fig. 3 shows $q/q_{\rm {0}}$ integrated over photon energy as a 
function of arc length assuming $q_{\rm {0}}$ appropriate to a
1.4$M_{\sun}$ object so that $L_{\rm {Edd}}$ is $\rm {1.76\times
10^{38}}$ erg s$^{-1}$. For the 4 cases, the neutron star luminosities are
$\rm {1.76\times 10^{36}}$, $\rm {7.04\times 10^{36}}$,
$\rm {3.52\times 10^{37}}$ and $\rm {1.67\times 10^{38}}$ erg s$^{-1}$.

Clearly, at low luminosities the X-ray emission peaks in a narrow strip
relatively close to the orbital plane, but extending down to $\theta$ = 0.
At high luminosities the hot region expands to fill the star. 
Inogamov \& Sunyaev showed that the calculated fluxes were not
sensitive to the magnitude of the viscosity parameter, as might be
expected.  
X-ray observations are not capable, of course, of
revealing the values $R\theta_{\rm {1}}$ and $R\theta_{\rm {2}}$
between which most of the X-ray emission is concentrated
assuming the model to be correct, but they do provide values of the
emitting area from fitting a simple blackbody. From the
measured luminosity and temperature, an area can be derived
and compared with the emitting area of the neutron star in the 
Inogamov \& Sunyaev theory, as described in the next section.

\section{Results}
\begin{table} 			
\caption{Emitting region on the surface of the neutron star calculated
using the theory of Inogamov \& Sunyaev (1999) expressed in terms of the range of
$R\,\theta$ and also in terms of half-heights $h$ derived for these
for 4 luminosities of the spreading layer $L_{\rm SL}$}
\begin{center}
\begin{tabular}{lrrrrr}
\hline\noalign{\smallskip}
$\;\;$$L_{\rm SL}/L_{\rm {Edd}}$&$\;\;$ $\Delta R\theta
$&$h$$\;\;$&$\Delta R\theta $&$h$$\;\;$ \\
        &                     &FWHM  &&$I_{\rm 0.5}^{\rm 0.99}$\\
        \%                     &km&km&km&km\\
\noalign{\smallskip\hrule\smallskip}
1  &0.12 &0.12  &0.09  &0.09 \\
4  &0.44 &0.44  &0.32  &0.32 \\
20 &2.47 &2.45  &1.51  &1.50 \\
80 &15.81&11.62 &9.14  &8.28 \\
\noalign{\smallskip}\hline
\end{tabular}
\end{center}
\end{table}

Fig. 3 shows the variation of flux with polar angle
$q(R\theta)$ expressed as a fraction of the Eddington flux $q_{\rm 0}$
in the theory of Inogamov \& Sunyaev
which we wish to compare with the results of the {\it ASCA} survey of LMXB.
Although Inogamov \& Sunyaev use an electron scattering dominated opacity for the
surface layers of the neutron star, they in fact, implement their model
assuming simple blackbody emission. As discussed in Sect. 1, whether modification
of the blackbody takes place depends on the electron density, and no evidence 
for modification in non-burst emission is known to exist.
Thus we simply obtain the theoretical sizes of the hot regions on the 
neutron star from Fig. 3 and compare these with blackbody areas from the survey.

We adopt two ways of defining the emitting area from the theoretical
variation $q(R\theta )$; comparison of these 
will indicate the uncertainty in the area.
Firstly, we use the region on the neutron star corresponding to the
FWHM of the flux $q(R\,\theta)$, i.e. the range between points where
$q$ = $q_{\rm {max}}/2$. Secondly, we used a definition
based on the intensity $I$ $\propto$ $\int\, q(\theta^{'})\,{\rm
cos}\theta^{'}\,d\theta^{'}$ integrated over the surface of the sphere
allowing for the angular dependence of the area $2\,\pi\,R^{\rm 2}\,{\rm
cos}\theta\,d\theta$. The intensity $I(\theta)$ increases asymptotically to 
a constant value corresponding to the maximum value of $R\,\theta$
at which there is X-ray emission. The range of $R\,\theta$ was then obtained
between $I$ = 0.50$\,I_{\rm max}$ and 0.99$\,I_{\rm max}$.
For each method, we calculated the half-height $h$ of an equatorial strip 
having the same area, from the range of $R\,\theta$ values using 
$h$ = $R\,{\rm sin}\theta$ with 12 km for the radius of the neutron star 
as assumed by Inogamov \& Sunyaev. These data are shown in Table 1.
The case of the highest luminosity point is interesting.
The FWHM in this case gives a value of $\Delta
(R\,\theta)$ of 15.8 km which corresponds to $\Delta \theta$ = 
75.5$\degmark $ if we assume $R$ = 12 km, and
90.5$\degmark $ for $R$ = 10 km. Thus, in the latter case, the whole star is clearly
emitting and the value of $h$ equal to $R{\rm sin}\theta$ becomes 10 km.
Alternatively, if we assume that 15.8 km is one quarter of the
circumference of the star, a radius of 10.1 km for  the neutron star is obtained.

\begin{figure}[!h]                                             
\begin{center}
\includegraphics[width=66mm,angle=270]{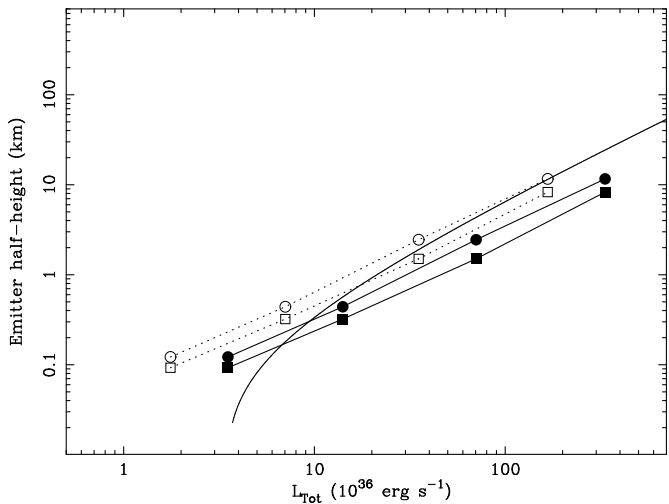}                     
\caption{Comparison of 4 theoretical points for the vertical
extent of the emitting region on the neutron star with $H(r_{\rm
{10}})$ (solid curve),
the height of the disk where $p_{\rm {r}}$ = 10$p_{\rm
{g}}$. Open squares and open circles are uncorrected (see text);
the squares show the height equivalent to 99\% of intensity,
the circles show the height equivalent to the FWHM of the flux
distribution. Filled squares and filled circles are
at corrected values of the total luminosity}
\end{center}
\end{figure}

In Fig. 4 we compare values of $h$
calculated in these two ways with $H(r_{\rm {10}})$, 
the height of the inner disk, as a function of total luminosity.
This function was shown to agree well with the {\it ASCA} survey
results by Church \& Ba\l uci\'nska-Church (2001),
and was calculated assuming a neutron star radius of 10 km.
Open squares
show the 99\% intensity points, and open circles show the FWHM
points. 
It can be seen that there is less than a factor of two
difference between the theoretical points of Inogamov \& Sunyaev
for the two methods of area calculation.
However, the contribution of the neutron star blackbody 
is only part of the total luminosity in X-ray binaries and so
the theoretical points should be corrected.
The other major emission component is that arising originally
as thermal emission in the accretion disk, which is then Comptonized
in the ADC (e.g. Church 2001; Ba\l uci\' nska-Church et al.
2001) making a dominant contribution in most LMXB. 
As Inogamov \& Sunyaev (1999) assumed the Newtonian approximation
in which the energy available at the inner edge of the accretion disk is one
half of the total energy, one possible method of correction would be to
plot the points at $L_{\rm Tot}$ = 2$\cdot L_{\rm BB}$, where $L_{\rm BB}$
is the luminosity of the spreading layer. The points corrected in this way are shown as
filled circles and squares.
The differing values of 
neutron star radius used by Inogamov \& Sunyaev (12 km) and in
the calculation of $H(r_{\rm {10}})$ (10 km) will have only 10\% 
effect on the comparison. 

\begin{figure}[!h]                                             
\begin{center}
\includegraphics[width=66mm,angle=270]{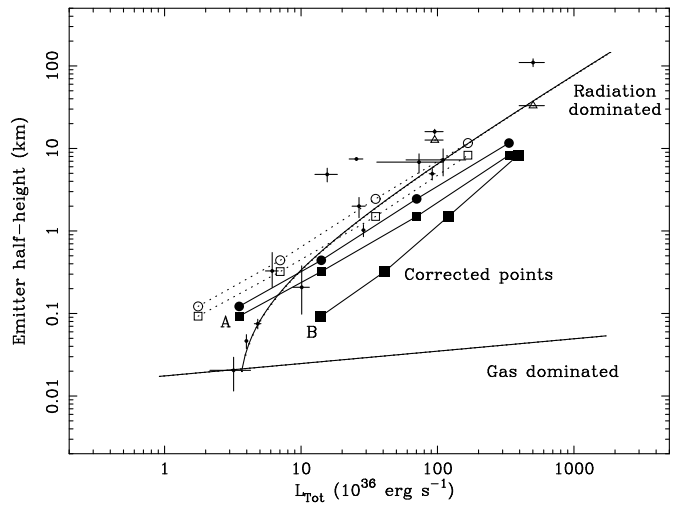}                     
\caption{Same as Fig. 4 with data points from the LMXB survey added
for luminosities in the band 1--30 keV.
The set of points labelled A are corrected as in Fig. 4
The points labelled B are corrected in luminosity based on
observational information from the LMXB survey. Also shown
is the variation of the thin disk half-height calculated close to
the stellar surface (at 1.001$\, R$) with $L$.
The theoretical height of the radiatively-supported disk (curved line)
approaches the thin disk line at low $L$ where
radiative support fails}
\end{center}
\end{figure}

In Fig. 5 we add the survey data points to the comparison.
The theoretical points are shown as before, uncorrected and also corrected by
a factor of two shift in $L_{\rm Tot}$. We now consider the extent of the
required correction further. 
Inogamov \&
Sunyaev (1999) assume that the spin of the neutron star is essentially
zero, and Sibgatullin \& Sunyaev (1998) argue that in this case,
$L_{\rm BB}$ = $2.8\cdot L_{\rm Tot}$. This implies that 
$L_{\rm Tot}$ = $1.34\cdot L_{\rm BB}$, so that we should correct the
theoretical points in Figs. 3 and 4 by moving them to higher values
of $L_{\rm Tot}$ by 34\%. However, this argument predicts that the
spectra of LMXB in general would be strongly dominated by the thermal
emission of the star, whereas, in fact, in many sources the
Comptonized emission strongly dominates, and work on the dipping
sources has proven that this originates in an extended ADC. It has been suggested,
for example by Inogamov \& Sunyaev (1999) that modification of the
neutron star emission in the stellar atmosphere will take place; however
this will not lead to a power law spectrum as observed,
but to a spectrum very similar to a simple blackbody (Madej 1991) and
the observed dominance of Comptonized emission in the spectrum could
only be avoided by assuming that most LMXB have a neutron star spinning
sufficiently fast that the emission due to viscous dissipation in the
surface is markedly reduced.
As an alternative to seeking a theoretical relation between $L_{\rm
Tot}$ and $L_{\rm BB}$, we can use an approximate observational relationship
provided by the {\it ASCA} survey. Fig. 1 shows that
there is not a simple relationship between 
$L_{\rm BB}$ and  $L_{\rm Tot}$, but rather that the blackbody
luminosities occupy a broad band of the two-dimensional space. To
make an approximate correction, we use the centre of this band.
This requires a substantial
movement of the theoretical points in Fig. 5 to higher values of $L_{\rm Tot}$
as shown by the filled squares labelled {\it B}.

However, we can avoid the difficulty of the correction
by comparing the theoretical points with the survey data replotted in
terms of $L_{\rm BB}$ instead of $L_{\rm Tot}$ as shown in Fig. 6.
\begin{figure}[!h]                                             
\begin{center}
\includegraphics[width=66mm,angle=270]{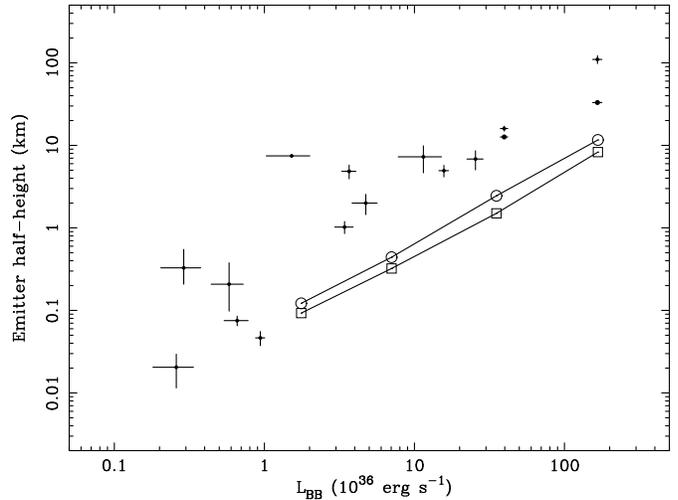}                     
\caption{The 4 theoretical points plotted as a fuction of blackbody luminosity
compared with data from the survey in the band 1--30 keV}
\end{center}
\end{figure}
A curve showing the theoretical height of the accretion disk is
not shown as this is a function of $L_{\rm Tot}$. It can be seen that
the survey points tend to converge with the theory at the lowest
luminosity, but are substantially larger than the theoretical points
at higher luminosities.

\section{Discussion}

The main result of the {\it ASCA} LMXB spectral survey was the agreement 
of $h$ = $H$, for sources in the quiescent state (i.e non-flaring).
However, it is not clear which of two explanations of this is correct:
that of radial flow between inner disk edge and star, in which the height 
of the inner radiatively supported disk {\it directly} determines $h$,
or the theory of Inogamov \& Sunyaev (1999) in which 
spreading of material on the neutron star surface to height $h$ depends on
the  mass accretion rate of which $H$ is a measure.
In the case of the first mechanism, the inner disk may have the
steep radial profile implied by standard theory close to the neutron
star, but radial flow takes place across the gap between inner disk and
star thus defining the emitting region on the star of similar height
to the highest point of the inner disk. Although there has been
extensive theoretical investigation of advective flow in black hole
systems (e.g. Abramowicz et al. 1996), very little work has been carried
out for neutron star binaries, and it is not clear whether the sonic
point lies within the stellar surface or not. However, Popham \& Sunyaev
(2001) show that the radial velocity increases by two orders of
magnitude within the disk. Two- or three-dimensional modelling would be 
required
to decide whether radial flow across the gap can take place.
A possible objection to the first mechanism is that the radiatively
supported inner disk may not be stable, i.e. implying that it cannot
exist. The stability of
the inner disk in a Galactic black hole system GRS\th 1915+109 has been
considered theoretically by Janiuk et al. (2000) who showed that oscillations
are expected leading to periodic variations in the X-ray luminosity of the 
system
similar to those observed. However, it was not found that this leads
to major disruption of the disk. Since the timescale for variability
depends on mass (Czerny, priv. comm.), we have applied this model to 
LMXB by scaling the stellar masses, and find a 
timescale of $\sim$ 30~s. Variability on this timescale is not
obviously observed, and would not in any case prove that major
disruption of the disk takes place.

In the present work, we have compared the survey results with the theory
of Inogamov \& Sunyaev (1999) in two ways. Firstly, we made an approximate
comparison as shown in Figs. 4 and 5. 
When a more appropriate comparison is made by comparing with $L_{\rm
BB}$, not $L_{\rm Tot}$, as shown in Fig. 6, the survey results and theory 
converge at the lowest luminosities, but at higher luminosities, the data,
i.e. the emitting areas, exceed the theory by a factor that we estimate
as between four and eight. However, the methods used of obtaining the emitting area
from theory underestimates the area, for example by using the FWHM, by up to
a factor of two. It consequently appears that the data exceed the theory by a factor of two to four.
Thus at low luminosities, the emitting area on the neutron star may be determined 
by accretion spreading, although even here radial flow would be an alternative 
explanation. At higher luminosities, the area exceeds the spreading theory
implying that the area is determined by radial flow which 
leads to the observed agreement between $h$ and $H$ and which enhances
the contribution due to spreading.
Although there is no observational evidence for any modification of
the neutron star blackbody spectrum in the non-burst emission of a LMXB,
this would lead to an underestimation of the emitting area, and
the disparity between the survey results and spreading theory would be 
somewhat increased.

Extension of the present accretion flow spreading theory which is 
essentially one-dimensional, to a two-dimensional form would be desirable. Further
analysis of LMXB spectra is clearly needed to increase the size of the
sample and to make more use of {\it BeppoSAX} data to constrain spectral 
fitting parameters better. However, the present results indicate that while
the neutron star accretion flow spreading theory of Inogamov \& Sunyaev
may set a {\it base level} to the neutron star blackbody emission, 
as observed in the faintest sources, it
underestimates the observed blackbody luminosity in brighter sources.
This implies that radial flow dominates leading to 
the agreement between $h$ and $H$.


\end{document}